# Contribution of Cosmic Rays from Sources with a Monoenergetic Proton Spectrum to the Extragalactic Diffuse Gamma-Ray Emission

*A.  Uryson*

Lebedev Physical Institute of Russian Academy of Sciences

Leninskii pr. 53, Moscow, 119333 Russia

**Abstract**—The extragalactic sources of ultra-high-energy ($E > 4 \times 10^{19}$ eV) cosmic rays that make a small contribution to the flux of particles recorded by ground-based arrays are discussed. We show that cosmic rays from such sources can produce a noticeable diffuse gamma-ray flux in intergalactic space compared to the data obtained with Fermi LAT (onboard the Fermi space observatory). A possible type of active galactic nuclei (AGNs) in which cosmic-ray protons can be accelerated to energies $10^{21}$ eV is considered as an illustration of such sources. We conclude that ultra-high-energy cosmic rays from the AGNs being discussed can contribute significantly to the extragalactic diffuse gamma-ray emission. In addition, a constraint on the fraction of the AGNs under consideration relative to the BL Lac objects and radio galaxies has been obtained from a comparison with the Fermi LAT data.

## INTRODUCTION

The sources of ultra-high-energy cosmic rays (UHECRs), $E > 4 \times 10^{19}$ eV, have not yet been established. At present, it is believed that UHECRs are extragalactic in origin and are emitted by active galactic nuclei (AGNs), but it is not known what AGNs these are.

UHECRs are investigated with the ground-based Pierre Auger Observatory (PAO) and Telescope Array (TA). The information from these arrays include the arrival directions of incident particles, their energy and mass composition - protons or nuclei.

The identification of sources by particle arrival directions turned out to be unsuccessful. The identification is complicated mainly by two factors. First, it is carried out by assuming that UHECRs propagate in intergalactic space almost rectilinearly. However, particles are apparently deflected by intergalactic magnetic fields. Second, the error in the particle arrival direction is $\sim 1^0$, so that quite a few objects fall into the region around the arrival direction, among which it is difficult to distinguish the CR source.

In intergalactic space UHECRs interact with background emissions. As a result of their interactions with the cosmic microwave background, the particle energy spectrum is "cut off", i.e., there is a lack of particles with energies $E > 10^{20}$ eV if CRs fly from distances exceeding $\sim 100$

Mpc (the GZK effect: Greisen 1966; Zatsepin and Kuzmin 1966). The PAO and TA energy spectra are cut off. However, they do not agree in shape at $E \geq 5 \times 10^{19}$ eV, and this difference is interpreted as a different UHECR mass composition: according to the PAO and TA data, these are nuclei and protons, respectively.

Apart from the GZK effect, the CR interaction with background emissions gives rise to electromagnetic cascades in intergalactic space (Hayakawa 1966; Prilutsky and Rozental 1970). Gamma-ray emission is produced in cascades, and it is measured within the extragalactic diffuse emission by Fermi LAT (the Large Area Telescope onboard the Fermi space observatory) (Ackermann et al. 2015).

Therefore, CRs are currently investigated by using not only their energy spectra, but also the cascade gamma-ray emission data. The UHECR models under consideration must now satisfy two criteria. First, it is required that the calculated UHECR energy spectra describe the measured spectrum. Second, the model intensity of the cascade gamma-ray emission must be lower than the measured intensity of the extragalactic diffuse emission minus the contribution of individual unresolved gamma-ray sources. Based on this scheme, the CR data were analyzed, for example, by Giacinti et al. (2015), Berezinsky et al. (2016), and Gavish and Eichler (2016). The first two papers are devoted to the UHECR composition (only protons or proton and nuclei); the second paper is devoted to the investigation of dark matter models, because the decays of dark matter particles contribute to the diffuse gamma-ray emission.

In this paper we show that there may exist UHECR sources that make a minor contribution to the flux of particles recorded by ground-based arrays. However, despite the insignificant intensity of particles on Earth, CRs from these sources can produce a noticeable diffuse gamma-ray flux in intergalactic space compared to the Fermi LAT data. This should be taken into account while analyzing CRs and dark matter models, because both the cascade emission from UHECRs and the decays of dark matter particles contribute to the diffuse emission.

As an illustration of the sources being discussed, we consider a possible type of AGN in which CR protons are accelerated to UHE near supermassive black holes.

Particles can be accelerated, for example, through the operation of the Blandford–Znajek mechanism (Blandford and Znajek 1977). A source power of ~$10^{46}$ erg s$^{-1}$ and a magnetic field maintained by a steady accreting plasma flow of $10^4$ G are sufficient for CRs to be accelerated to UHE in this mechanism without invoking any exotic scenarios.

Particles can also be accelerated by an electric field in the accretion disk regions where an explosive field growth occurs (Haswell et al. 1992), while in more exotic cases particle acceleration to $10^{21}$ eV is possible if a supermassive black hole is surrounded by a superstrong magnetic field (Kardashev 1995; Shatsky and Kardashev 2002; Zakharov et al. 2003).

The computations were performed with the TransportCR code (Kalashev and Kido 2015).

## THE MODEL

Electromagnetic cascades result mainly from the following reactions. In intergalactic space UHECRs interact with microwave and radio emissions $p + \gamma_{rel} \rightarrow p + \pi^0$ and $p + \gamma_{rel} \rightarrow n + \pi^+$. The decays of the forming pions $\pi^0 \rightarrow \gamma + \gamma$, $\pi^+ \rightarrow \mu^+ + \nu_\mu$ and muons $\mu^+ \rightarrow e^+ + \nu_e + \nu_\mu$ give rise to positrons and gamma-ray photons, while they generate electromagnetic cascades in the reactions with microwave emission and extragalactic background light $\gamma + \gamma_b \rightarrow e^+ + e^-$ (pair production) and $e + \gamma_b \rightarrow e^- + \gamma^-$ (inverse Compton effect).

The assumptions adopted in the model refer to three items: the injection spectra and cosmic evolution of UHECR sources, extragalactic background emissions, and extragalactic magnetic fields.

We also assume that the UHECR sources are point-like. These are AGNs where charged particles are accelerated to energy $10^{21}$ eV near supermassive black holes by electric fields (Haswell et al. 1992; Kardashev 1995; Shatsky and Kardashev 2002). Because of these acceleration mechanisms, we assume that the CR injection spectrum is monoenergetic with energy $E=10^{21}$ eV.

We also assume that UHECRs consist of protons.

The cosmic evolution of the objects under consideration is associated with the evolution of the states of supermassive black holes (see, e.g., Kardashev 1995). It is unclear and therefore we consider two possible AGN evolution scenarios: as in BL Lac objects (Di Mauro et al. 2014; Kalashev and Kido 2015) or stronger and as in radio galaxies (Smolčic et al. 2017). In the case of evolution as in radio galaxies, we took into account the evolution of only the density of objects, because it is unclear how their luminosity is related to the redshift.

The extragalactic background emissions were treated as follows.

The cosmic microwave background radiation has a Planck energy distribution with a mean $\varepsilon_r = 6.7 \times 10^{-4}$ eV. The mean photon density is $n_r = 400$ cm$^{-3}$.

The characteristics of the extragalactic background light were taken from (Inoue et al. 2013).

To describe the background radio emission, we used the model of the luminosity evolution for radio galaxies from (Protheroe and Biermann 1996).

The magnetic field in intergalactic space is apparently nonuniform (Kronberg 2005; Essey et al. 2011; Dzhatdoev et al. 2017): there are regions where the magnetic field is $1 \times 10^{-17}$ G $< B < 3 \times 10^{-14}$ G and filamentary areas in which the magnetic field is stronger: $B \approx 10^{-9}$ - $10^{-8}$ G. In both cases, the cascade electrons lose their energy through synchrotron radiation insignificantly (Uryson 2017a).

## RESULTS

The UHECR energy spectra calculated for two source evolution scenarios are shown in Fig. 1. This figure also shows the PAO spectrum fit by Verzi et al. (2017).

In both cases of the evolution of sources, the model CR spectra are lower than the PAO spectrum by several orders of magnitude.

In addition, the calculated spectra differ greatly from the PAO spectrum in shape (they differ equally greatly from the TA UHECR spectrum).

We compare the model spectra with the PAO measurements although the proton composition of UHECRs is assumed in the adopted model, while according to the PAO data these are nuclei. The reason for the such comparison is as follows. The model spectra are normalized to the PAO spectrum at an energy of $10^{19.5}$ eV ($3.16 \times 10^{19}$ eV). At this and lower energies the difference in the TA and PAO spectra is small: 20–30%. At $E \geq 5 \times 10^{19}$ eV the intensities in the measured spectra differ: the intensity from the PAO data is lower than that from the TA data by a factor of 8–9 (Verzi et al. 2017). Therefore, the model CR spectra are lower than both PAO and TA spectra by several orders of magnitude.

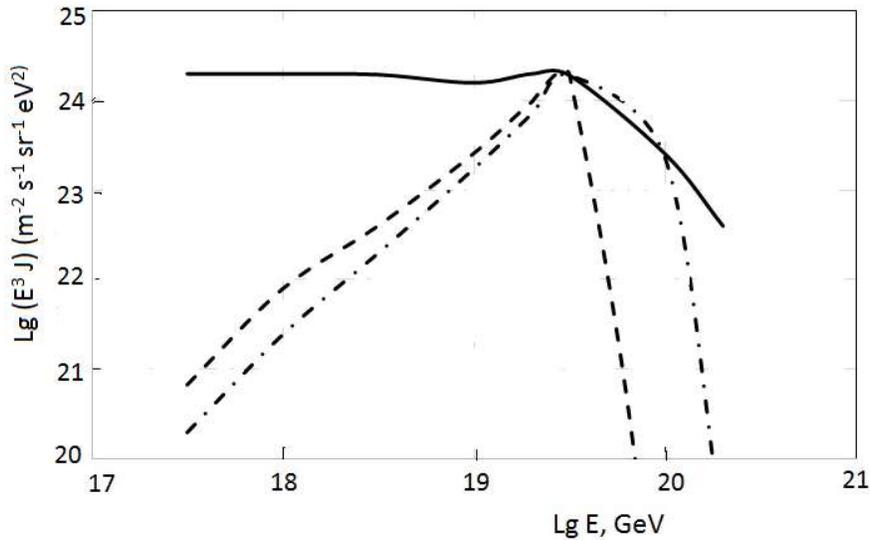

**Fig. 1.** PAO UHECR energy spectrum, its fit (Verzi et al. 2017) (solid line), and model CR spectra: for the evolution of sources as in radio galaxies (dashed line) and for the evolution as in BL Lac objects (dash–dotted line). The model spectra are normalized to the PAO spectrum at the energy of $10^{19.5}$ eV.

Let us discuss the intensity of the gamma-ray emission that UHECRs initiate in intergalactic space.

The cascade gamma-ray emission spectra are analyzed in detail in (Berezinsky and Kalashev, 2016). In the latter paper the shape of the spectrum was found to be virtually independent of the initial spectrum of the particles that initiated the cascade. The goal of our paper is to compare the gamma-ray emission obtained in the adopted model with the Fermi LAT measurements. For such a comparison we use the integrated intensity of the cascade emission, as is done in (Berezinsky et al. 2016). Therefore, here we do not discuss the model cascade gamma-ray emission spectra.

The integrated intensity of the cascade gamma-ray emission at energies $E > 50$ GeV was found from the differential intensity of the cascade gamma-ray emission calculated with the TransportCR code:

$$I\gamma(E > 50 \text{ GeV, evolution of sources: BL Lac objects}) = 1.002 \times 10^{-9} \text{ (cm}^{-2}\text{ s}^{-1}\text{ sr}^{-1}), \quad (1)$$

$$I\gamma(E > 50 \text{ GeV, evolution of sources: radio galaxies}) = 1.641 \times 10^{-9} \text{ (cm}^{-2}\text{ s}^{-1}\text{ sr}^{-1}). \quad (2)$$

The energy range $E > 50$ GeV is chosen here because the contribution of individual unresolved gamma-ray sources was obtained for it (Di Mauro 2016). We take into account this contribution below when comparing the intensity of the cascade gamma-ray emission with the Fermi LAT data. The difference in integrated intensity (1) and (2) is attributable to different source evolution scenarios.

## DISCUSSION

Let us compare the model integrated intensity of the gamma-ray emission with the Fermi LAT data (Ackermann et al. 2015). The extragalactic isotropic diffuse gamma-ray background (IGRB) measured with Fermi LAT is

$$\text{IGRB}(E > 50 \text{ GeV}) = 1.325 \times 10^{-9} \text{ (cm}^{-2}\text{ s}^{-1}\text{ sr}^{-1}). \quad (3)$$

This includes the emission from individual unresolved gamma-ray sources. At energies above 50 GeV their contribution is 86 $(-14, +16)$% (Di Mauro 2016).

Excluding the contribution of unresolved sources, 86%, from the IGRB, we obtain

$$\text{IGRB}_{\text{without blazars}}(E > 50 \text{ GeV}) = 1.85 \times 10^{-10} \text{ (cm}^{-2}\text{ s}^{-1}\text{ sr}^{-1}). \quad (4)$$

This value is much lower than the calculated intensity of the cascade gamma-ray emission (Eqs. (1) and (2)).

Taking into account the measurement errors and the uncertainty in the galactic diffuse emission models that are used to derive the IGRB from the Fermi LAT data, and assuming that given the error of 14%, the contribution of unresolved gamma-ray sources is 72% (rather than 86%, as is adopted in the estimate (4)), we obtain the following band of IGRB$_{\text{without blazars}}$ ($E > 50$ GeV):

$$2.20 \times 10^{-10} \text{ (cm}^{-2}\text{ s}^{-1}\text{ sr}^{-1}) \leq \text{IGRB}_{\text{without blazars}}(E > 50 \text{ GeV}) \leq 5.40 \times 10^{-10} \text{ (cm}^{-2}\text{ s}^{-1}\text{ sr}^{-1}). \quad (5)$$

The calculated intensity of the cascade gamma-ray emission here is also several times higher than the measured extragalactic diffuse gamma-ray emission.

The reason for the excess is that the calculated UHECR flux was normalized to the measured flux at an energy of $10^{19.5}$ eV. However, the AGNs under consideration are not the main UHECR sources and their contribution to the CR flux is definitely smaller. Therefore, from the condition

$$I\gamma(E > 50 \text{ GeV}) < \text{IGRB}_{\text{without blazars}}(E > 50 \text{ GeV}), \qquad (6)$$

where we take into account that the main UHECR sources and the decays of dark matter particles also contribute to the IGRB, we find the fraction $R$ of the AGNs under consideration among the BL Lac objects and radio galaxies:

$$R < 18\% \text{ in comparison with BL Lac objects,}$$

$$R < 11\% \text{ in comparison with radio galaxies.}$$

As an illustration, consider an example where the fraction of the AGNs under consideration is 10% of the BL Lac objects and radio galaxies. The model intensity of the cascade gamma-ray emission is then

$I\gamma$ ($E > 50$ GeV, evolution of sources: BL Lac objects) =

$$1.002 \times 10^{-10} \text{ (cm}^{-2} \text{ s}^{-1} \text{ sr}^{-1}) = 0.54 \cdot \text{IGRB}_{\text{without blazars,}} \qquad (7)$$

$I\gamma$ ($E > 50$ GeV, evolution of sources: radio galaxies) =

$$1.641 \times 10^{-10} \text{ (cm}^{-2} \text{ s}^{-1} \text{ sr}^{-1}) = 0.89 \cdot \text{IGRB}_{\text{without blazars.}} \qquad (8)$$

These values fall into the IGRB band (5) and the possibility remains for the contribution of UHECRs from the main sources to the extragalactic diffuse gamma-ray emission and for a possible contribution of dark matter particle decays.

In the model under consideration the particles have a low flux on the Earth, but they generate a significant cascade gamma-ray emission. This is related mainly to the model proton injection spectrum: it is monoenergetic, all particles are emitted with energy $E = 10^{21}$ eV. The mean free paths of particles with such an energy is $\sim 10$ Mpc and, as a result, the CR energy is efficiently transferred to the cascade and, consequently, to the gamma-ray emission. A detailed analysis is provided in (Uryson 2017b). It follows from (7) and (8) that stronger evolution of sources (as in radio galaxies) leads to an increase in the intensity of the cascade gamma-ray emission.

Here we do not analyze the cascade gamma-ray emission spectrum, because its shape does not depend on the UHECR injection spectrum (Berezinsky et al. 2016).

## CONCLUSION

We considered the possibility of the existence of extragalactic UHECR sources that make a minor contribution to the flux of particles recorded by ground-based arrays. As an illustration of such UHECR sources, we analyzed a possible type of AGN in which CRs are accelerated to energies of $10^{21}$ eV near supermassive black holes. Particles can be accelerated in electric fields (Haswell et al. 1992; Kardashev 1995; Shatsky and Kardashev 2002). Due to these mechanisms, we assumed the CR injection spectrum to be monoenergetic with energy $E = 10^{21}$ eV.

It has been found that the particle flux from these sources is considerably lower than the flux measured at PAO and TA. However, UHECRs can produce a noticeable diffuse gamma-ray flux in intergalactic space compared to the Fermi LAT data.

We draw the following conclusions from the results obtained.

There may exist UHECR sources that give a low particle flux on Earth, but the emitted CRs can contribute significantly to the extragalactic diffuse gamma-ray emission. This should be taken into account analyzing dark matter models and UHECR data.

It follows from a comparison with the Fermi LAT data that in the model adopted the fraction of AGNs where particles are accelerated near supermassive black holes having nearly monoenergetic injection spectra, is: $R < 18\%$ compared to BL Lac objects with evolution according to (Di Mauro et al. 2014) and $R < 11\%$ compared to radio galaxies in which the density evolution is described in (Smolčic et al. 2017).

The results obtained depend on the contribution of individual unresolved gamma-ray sources to the extragalactic diffuse emission. At present, it has been determined with an error of about 15% (Di Mauro 2016). This contribution can be refined using instruments with a better angular resolution than that of Fermi LAT. For example, for the GAMMA-400 space instrument the angular resolution at an energy of 100 GeV is $\sim 0.01^0$ (Topchiev et al. 2017), while for Fermi LAT the angular resolution at the same energy is $0.05$–$0.1^0$ (Fermi Lat Performance).

### Acknowledgments

I thank O.E. Kalashev for the discussion of the code TransportCR and N.P. Topchiev for the discussion of gamma-ray telescope characteristics. I am also grateful to the referees for the discussion and remarks.